\documentstyle[aps,eqsecnum,twoside]{revtex}
\newcommand{\C}{{\if mm {{\rm C}\mkern -15mu{\phantom{\rm t}\vrule}}
\mkern +10mu \else \leavemode \hbox{I}\kern -.17em \hbox{C} \fi}}
\newcommand {\bG}{\bar {\Gamma}}
\newcommand {\cM}{\cal M}
\newcommand {\bA}{\bf A}
\newcommand {\G}{\Gamma}
\newcommand {\vf}{\varphi}
\newcommand {\vt}{\vartheta}
\newcommand {\ve}{\varepsilon}
\newcommand {\bV}{\bar V}
\newcommand {\ba}{\bar \alpha}
\newcommand {\bb}{\bar \beta}
\newcommand {\bs}{\bar \sigma}
\newcommand {\bt}{\bar \tau}
\newcommand {\bl}{\bar \lambda}
\newcommand {\be}{\bar \varepsilon}
\newcommand {\p}{\partial}
\newcommand {\pr}{\prime}
\begin{document}
\title{On the Natural Gauge Fields of Manifolds}
\author{A. B. Pestov\\
Bogoliubov Laboratoty of Theoretical Physics\\
Jonit Institute for Nuclear Research\\
141980 Dubna, Moscow region, Russia\\
e-mail: pestov@thsun1.jinr.ru}
\author{Bijan Saha\\ 
Laboratoty of Information Technologies\\
Jonit Institute for Nuclear Research\\
141980 Dubna, Moscow region, Russia\\
e-mail: bijan@cv.jinr.ru, saha@thsun1.jinr.ru}
\date{}
\maketitle
\thispagestyle{empty}
\begin{abstract}
The gauge symmetry inherent in the concept of manifold has been discussed.
Within the scope of this symmetry the linear connection or
displacement field can be considered as a natural gauge field on the
manifold. The gauge invariant equations for the displacement field
have been derived. It has been shown that the energy-momentum tensor
of this field conserves and hence the displacement field can be
treated as one that transports energy and gravitates. To show the
existence of the solutions of the field equations we have derived the
general form of the displacement field in Minkowski space-time which
is invariant under rotation and space and time inversion.  With this
anzats we found spherically-symmetric solutions of the equations in
question.

\end{abstract}

{\bf PACS} number(s): 04.20, 11.15, 02.30.H\\

\vspace*{1cm}
\section{Introduction}
According to the modern standpoint, space-time theory is the one that
possesses a mathematical representation whose elements are a smooth
four-dimensional manifold $\cM$ and geometric objects defined on
this manifold. The geometry on the manifold is defined by metric
and linear connection. In general, the linear connection is in no way
related to the metric since these concepts define on the manifold
$\cM$ different geometric operations. The metric on the manifold defines
the length of a curve while the linear connection defines parallel
transport (displacement) of vectors along arbitrary path on $\cM$.
It should be emphasized that soon after the creation of General Relativity
A. Eddington put forward the idea to derive all theory on the basis of
parallel displacement only \cite{Eddington}. Here the metric and the linear 
connection as a totally independent geometric objects by structure will be 
considered as fundamental fields. 
It is our principal assumption. According to the
fundamental idea of Einstein, metric corresponds to gravitational field
while all other fields, being the source of gravitational one, carry energy.
Hence and from the above made assumption it follows that, like the
electromagnetic field, the field of parallel displacement carries energy
and appears to be the source of gravitational field, possessing geometric
meaning. Thus, our aim is to derive natural equations for the field of
parallel displacement and obtain the relevant conserving energy-momentum
tensor, i.e., to show that within the frame-work of the canonical
Einstein theory of gravity the linear connection can be considered
on the same level with electromagnetic one.

\section{Symmetry group}
There are two symmetry groups closely connected with concept of
manifold.  One of them is a group of transformations of the manifold
$\cM$ itself, the manifold mapping group, and the other is a group of
transformations acting in tangent vector spaces $T_p (\cM) $. The
latter concept is clearly expounded in the treatise by Misner,
Thorne, Wheeler ~\cite{Misner}. The well-known manifold mapping
group~\cite{Anderson} is often called the group of general
transformations of coordinates or the group of diffeomorphisms. The
physical meaning of the manifold mapping group is that it is a group
of symmetry of gravitational interactions in Einstein theory of
gravity.  A systematic and thorough consideration of the questions
connected with space-time symmetry of General Relativity may be found
in Ref.~\cite{Anderson}.  We emphasize only that the diffeomorphism
group is evidently the widest group of space-time symmetry.

Let the given vector field $V^i$ undergoes infinitesimal parallel 
displacement, then
\begin{eqnarray}
dV^i +\G^{\,\,i}_{jk} V^k d x^j = 0,
\label{pd}
\end{eqnarray}
where $\G_{jk}^{\,\,i}$ are the components of linear connection.
Vector fields form linear vector space $L$. The isomorphic mappings of
the vector space $L$ onto itself is defined by the tensor fields of
type $(1,1)$.
Let $S^i_j$ be the components of a tensor field $S$ of type $(1,1)$ that
satisfies the condition  $ \det (S^i_j) \ne 0 $ only.  In this case, there
exists a tensor field $S^{-1}$ with components $T^i_j$ such that
$S^i_k \,T^k_j = \delta^i_j.$ Now a tensor field $S$
can be regarded as an isomorphism of $L$ onto itself
\begin{equation}
\bV^i (x)=S^i_j (x) V^j (x).
\label{trans}
\end{equation}
Since there is no objective reasons to distinguish vector fields
$\bV^i (x)$ and $V^j (x)$,
we want to define the law of parallel displacement for the vector
$\bV^i$ induced by (\ref{pd}) and (\ref{trans}). It
can be shown that if the vector $V^i$ undergoes parallel displacement
(\ref{pd}) then $\bV^i$, defined by (\ref{trans}), undergoes parallel
displacement
{\rm }\begin{eqnarray}
d\bV^i + \bG^{\,\,i}_{jk} \bar V^k d x^j = 0,
\label{pdb}
\end{eqnarray}
where
\begin{eqnarray}
\bG^{\,\,i}_{jk}  = S^i_l\G^{\,\,l}_{jm}  T^m_k  + S^i_l\partial_j T^l_k,
\label{bg}
\end{eqnarray}
and $T^i_j$  are components of the field $S^{-1}$  inverse to $S.$
In what follows we will
consider the transformation group ~(\ref{trans}) as the natural
group of gauge transformation inherent in the manifold $\cM$. The
transformation~(\ref{bg}) give the realization of gauge group on the fields
of parallel displacement and in that sense $\G_{jk}^{\,\,i}$ are analogical
to the potentials of electromagnetic fields.
Now our aim is to find equation for $\G$ that is invariant under
(\ref{bg}). From ~(\ref{bg}) it follows that if $\G_{jk}^{\,\,i}$ are
the components of linear connection than $\bG_{jk}^{\,\,i}$ are too the
components of linear connection, that is, under coordinate
transformation $\bG$ transforms in accordance with the same well-known laws
as does it $\G$ itself \cite{Anderson}.

\section{ Gauge-Invariant Equations}

As it is noted above, the diffeomorphism group is responsible for
gravitational interactions, and thus, the gauge group under consideration
is a symmetry group of new interactions. To simplify
computations and to write equations in a symmetrical and manifestly
gauge - invariant form, we introduce the notion of the gauge
derivative. We will say that a tensor field $T$ of the type $(m,n)$ is
of the gauge type $(p,q)$ if under the transformations of the gauge
group there is the correspondence
\begin{displaymath}
T \Rightarrow
{\bar T} ={ \underbrace{S \cdots S}_p} T{\underbrace{ S^{-1} \cdots
S^{-1}}_q},
\end{displaymath}
where
\begin{displaymath}
0\leq  p{ \leq m} \mbox { and } 0 \leq q {\leq n}.
\end{displaymath}
The Einstein potentials $g_{ij}$ being a tensor field of the type $(0,2)$
is to be assigned the gauge type $(0,0)$ because the Einstein Equations are
not invariant with respect to the transformations
$\bar g_{ij} = g_{kl} T_{i}^{k} T_{j}^{l}$. Let the vector field $V^i$
has a gauge type $(1,0)$. We define gauge derivative as
$$ D_j V^k = \p_j V^k + \G_{jl}^{\,\,k} V^l$$
Now, if the equality
$$\bar D_j \bV^k = \p_j \bV^k + \bG_{jl}^{\,\,k} \bV^l$$
holds, then from ~(\ref{trans}) and ~(\ref{bg}) follows that
$$\bar D_i \bV^j = S_{k}^{j} D_i V^k.$$
Since
$$D_i D_j V^k = \p_i (D_j V^k) + \G_{il}^{\,\,k} \bigl(D_j V^l\bigr),$$
then $D_i D_j V^k$ is not a tensor field, nevertheless, the commutator
of gauge derivatives are tensor fields, as
\begin{equation}
[D_i,D_j] V^k = B_{ijl}^{\,\,\,\,k} V^l,
\label{com}
\end{equation}
where
\begin{equation}
B_{ijl}^{\,\,\,\,k} = \p_i \G_{jl}^{\,\,k} - \p_j \G_{il}^{\,\,k}
+ \G_{im}^{\,\,k} \G_{jl}^{\,\,m} - \G_{jm}^{\,\,k} \G_{il}^{\,\,m},
\label{rel}
\end{equation}
is the Riemann tensor of curvature of connection $\G$. Note that
$B_{ijl}^{\,\,\,\,k}$ is a tensor field of type $(1,3)$ and gauge
type $(1,1)$.
In what follows, for brevity, we use matrix notation, assuming that
$$\G_i = \bigl(\G_{ij}^{\,\,k}\bigr), \quad B_{ij} = \bigl(
B_{ijl}^{\,\,\,\,k}\bigr), \quad {\rm Tr} B_{ij} = B_{ijk}^{\,\,\,\,k},
\quad \G_i\,\G_j = \G_{im}^{\,\,k} \G_{jl}^{\,\,m}.$$
In matrix notation
\begin{eqnarray}
B_{ij} &=& \p_i \G_j - \p_j \G_j + [\G_i,\G_j], \label{H}\\
\nonumber\\
\bG_i &=& S \G_i S^{-1} + S \p_i S^{-1}, \label{bgm}\\
\nonumber\\
\bar B_{ij} &=& S B_{ij} S^{-1}. \label{bH}
\end{eqnarray}
It is obvious from (\ref{H}) - ~(\ref{bH}) that, like
$F_{ij} = \p_i \bA_j - \p_j \bA_i$, $B_{ij}$
is strength tensor. The generally covariant and gauge-invariant Lagrangian
for gauge field $\G_{ij}^{\,k}$ (displacement field) has the form
\begin{equation}
{\cal L} = -\frac{1}{4} {\rm Tr} \bigl(B^{ij} B_{ij}\bigr),
\label{lag}
\end{equation}
where
$$B^{ij} = g^{ik} g^{jl} B_{kl},$$
and $g^{ij}$ is a tensor field inverse to $g_{ij}$ such that
$g_{jk} g^{ik} = \delta_{j}^{i}.$

Varying ~(\ref{lag}) with respect to $\G$, we obtain the following
system of second order differential equations for the displacement field
\begin{equation}
\frac{1}{\sqrt{-g}} D_i \bigl(\sqrt{-g} B^{ij}\bigr) = 0.
\label{cl}
\end{equation}
In fact, if $\delta \G_i$ is variation, then
$$\delta B_{ij} = D_i \delta \G_j - D_j \delta \G_i.$$
Hence it follows that
$$\delta {\cal L} = - {\rm Tr} \bigl( B^{ij} D_i \delta \G_j\bigr) =
 - \p_i {\cal J}^i + {\rm Tr} \bigl( (D_i B^{ij}) \delta \G_j\bigr), $$
where ${\cal J}^i = {\rm Tr} (B^{ij} \delta \G_j).$ Since
$$\p_i {\cal J}^i = \frac{1}{\sqrt{-g}} \p_i \bigl(\sqrt{-g} {\cal J}^i
\bigr) - \Bigl(\frac{\p_i \sqrt{-g}}{\sqrt{-g}}\Bigr) {\cal J}^i,$$
then
$$\delta {\cal L} = -
\frac{1}{\sqrt{-g}} \p_i \bigl(\sqrt{-g} {\cal J}^i \bigr) + {\rm Tr}
\Bigl(\frac{1}{\sqrt{-g}} D_i \bigl(\sqrt{-g} B^{ij}\bigr)
\delta \G_j \Bigr). $$
Q.E.D. $\Box$ 

Varying the action
$${\cal A} = \int\, dx^4\,{\cal L} \sqrt{-g}$$
with respect to the metric $g_{ij}$ we obtain gauge-invariant
energy-momentum metric tensor for displacement field $\G$
\begin{equation}
T^{ij} = {\rm Tr} \Bigl(B^{ik} B_{\,\,k}^{j}\Bigr) - \frac{1}{4} g^{ij}
\Bigl(B_{kl} B^{kl}\Bigr)
\label{emt}
\end{equation}
which on the solutions of the equations ~(\ref{cl}) satisfies the local
law of energy conservation
\begin{equation}
T^{ij}\,_{\!;j} = 0.
\label{emtc}
\end{equation}
Here semicolon denotes the covariant derivative with respect to the
Levi - Civita connection belonging to the metric $g_{ij}$
\begin{eqnarray}
\{^{\,\,i}_{jk}\} =
\frac{1}{2} g^{il} (\p_j g_{kl} + \p_k g_{jl} - \p_l g_{jk} ).
\label{LC}
\end{eqnarray}
In view of its significance, we underline few details of the proof of 
the relation ~(\ref{emtc}). We have
\begin{eqnarray}
T^{ij}\,_{\!;j} = \p_j T^{ij} + \{^{\,\,j}_{jk}\} T^{ik} +
\{^{\,\,i}_{jk}\} T^{jk} =
\frac{1}{\sqrt{-g}} \p_j \bigl(\sqrt{-g} T^{ij} \bigr) +
\{^{\,\,i}_{jk}\} T^{jk}.
\label{tc}
\end{eqnarray}
Since, according to ~(\ref{LC})
\begin{equation}
\p_i g^{jk} = - \{^{\,\,j}_{il}\} g^{kl} - \{^{\,\,k}_{il}\} g^{jl},
\label{m}
\end{equation}
then it can be shown that
\begin{eqnarray}
{\rm Tr} \Bigl(B^{jk} D_j B^{i}_{\,\,k}\Bigr) &=&
-\{^{\,\,i}_{jk}\}{\rm Tr} \Bigl(B^{jl} B^{k}_{\,\,l}\Bigr)
-\{^{\,\,l}_{jk}\} g^{ik}{\rm Tr} \Bigl(B^{jm} B_{lm}\Bigr) + \frac{1}{2}
g^{ik} {\rm Tr} \Bigl( B^{jl} (D_j B_{kl} - D_l B_{kj})\Bigr)
\label{Tr1}\\
{\rm Tr} \Bigl(B_{kl} D_j B^{kl}\Bigr) &=&
{\rm Tr} \Bigl(B^{kl} D_j B_{kl}\Bigr) -
4 \{^{\,\,k}_{jl}\} {\rm Tr} \Bigl(B_{km} B^{lm}\Bigr) \label{Tr2}
\end{eqnarray}
From ~(\ref{tc}), (\ref{Tr1}) and (\ref{Tr2})  follows
\begin{eqnarray}
T^{ij}\,_{\!;j} = {\rm Tr}
\Bigl(\frac{1}{\sqrt{-g}} D_j \bigl(\sqrt{-g} B^{jk}\bigr) B^{i}_{\,\,k}
\Bigr)
+ \frac{1}{2} g^{ik} {\rm Tr} \Bigl(B^{jl} \bigl(
D_j B_{kl} + D_k B_{lj} + D_l B_{jk})\Bigr). \nonumber
\end{eqnarray}
Since the equation
\begin{eqnarray}
D_j B_{kl} + D_k B_{lj} + D_l B_{jk} = 0, \nonumber
\end{eqnarray}
is fulfilled identically, then the local law of energy-momentum conservation
(\ref{emtc}) is also fulfilled for the case in question.

Our conclusion is that the equations ~(\ref{cl}) and the Einstein equations
\begin{eqnarray}
R_{ij} - \frac{1}{2} g_{ij} R = \kappa T_{ij},
\label{EEQ}
\end{eqnarray}
with the right-hand side given by the expression ~(\ref{emt}), form
a consistent system of partial differential equations which is
invariant under gauge transformations as well as under the
transformations of diffeomorphism group. Now we have proved that the
displacement field $\G$ is really the origin of gravitational field
within the scope of given gauge approach.

\section{Spherical-symmetrical gauge potentials}

As the first step to investigate the equations~(\ref{cl}) it is very
important to show that they have non-trivial solutions. In doing this
we show that the equations (\ref{cl}) possess spherically symmetric
solutions.

The general theory of space-time symmetry within the scope of
theory of gauge fields has been developed in ~\cite{Manton1},
~\cite{Manton2}. We apply the results obtained there to our
particular case. Note that the spherically symmetric solutions of
$SU(2)$ Yang-Mills equations were first derived by Ikeda and Miyachi
\cite{Ikeda} and for $SU(3)$ by Loos \cite{Loos}.

In this section we consider Minkowski space-time with the metric
in spherical system of coordinates that is most convenient under the
consideration of $SO(3)$ symmetry:
\begin{equation}
ds^2 = dt^2 - dr^2 - r^2 d\vt^2 - r^2 {\rm sin}^2\,\vt d\vf^2,
\label{metric}
\end{equation}
where $c$ has been taken to be unity.

First of all we would like to
find gauge potentials those are invariant under the displacement along the
time axis $t\rightarrow t + a$, i.e.,
\begin{eqnarray}
\G ^i_{jk}(x^0, x^1, x^2, x^3) = \G ^i_{jk} (x^0 + a, x^1, x^2, x^3)
\label{td}
\end{eqnarray}
From (\ref{td}) follows that all $\G$'s  are independent of $t$.
Now we shall look for $S0(3)$ invariant gauge potentials.
Generators of $SO(3)$ group in spherical coordinates have
the form~\cite{Petrov}
\begin{mathletters}
\label{gen}
\begin{eqnarray}
X_1 &=& {\rm sin}\,\vf \frac{\p}{\p \vt} + {\rm cot}\,\vt\, {\rm cos}\,\vf
\frac{\p}{\p \vf}, \\
X_2 &=& -{\rm cos}\,\vf \frac{\p}{\p \vt} + {\rm cot}\,\vt\, {\rm sin}\,\vf
\frac{\p}{\p \vf}, \\
X_3 &=& - \frac{\p}{\p \vf},
\end{eqnarray}
\end{mathletters}
and hence, the problem is to find solutions of the equations $ L_X
\G = 0$ when Lie derivative are taken along the vectors fields of
$SO(3)$ group. The equation $ L_{X_a} \G = 0,\,\,\, a = 1,2,3$ can
be written in the following matrix representation
\begin{equation}
L_{X_{a}}
\G_j = V^\ell_{(a)} \p_{\ell} \G_j + [\G_j , A_{(a)}] + \G_{\ell}
A_{(a) j}^{\ell} + \p_j A_{(a)} = 0,
\label{LD}
\end{equation}
Here, $V_{(a)}^{\ell}$ are defined  from $X_{(a)} = V_{(a)}^{\ell}
\p_{\ell}$ as
\begin{eqnarray} V_{(1)}^{\ell} &=& (0,\, 0,\, {\rm
sin}\,\vf,\, {\rm cot}\,\vt\, {\rm cos}\,\vf),\nonumber\\
V_{(2)}^{\ell} &=& (0,\, 0,\, -{\rm cos}\,\vf,\, {\rm cot}\,\vt\,
{\rm sin}\,\vf),\nonumber\\
V_{(3)}^{\ell} &=& (0, 0, 0, -1). \nonumber
\end{eqnarray}
The matrices $A_{(a)}$ here take the forms
\begin{eqnarray}
A_{(1)} &=& A_{(1)j}^{i} = \p_j V_{(1)}^{i}
= \left(\begin{array}{cc}
0&0\\
0&\tilde{A}_{(1)}\end{array}\right), \quad
\tilde{A}_{(1)} = \left(\begin{array}{cc}
0&{\rm cos} \,\vf\\
-{\rm cos}\,\vf/{\rm sin}^2 \,\vt &-{\rm cot} \,\vt\, {\rm sin} \,\vf
\end{array}\right), \nonumber\\[2mm]
A_{(2)} &=& A_{(2)j}^{i} = \p_j V_{(2)}^{i}
= \left(\begin{array}{cc}
0&0\\
0&\tilde{A}_{(2)} \end{array}\right), \quad
\tilde{A}_{(2)} = \left(\begin{array}{cc}
0&{\rm sin} \,\vf\\
-{\rm sin} \,\vf/{\rm sin}^2 \,\vt &-{\rm cot} \,\vt\, {\rm cos} \,\vf
\end{array}\right), \nonumber\\[2mm]
A_{(3)} &=& A_{(3)j}^{i} = \p_j V_{3}^{i} \,=\, 0. \nonumber
\end{eqnarray}
Let us write the equations $L_{X_{(a)}} \G_j = 0$ explicitly. The
equations $L_{X_{(1)}} \G_j = 0$ can be written as follows
\begin{mathletters}
\label{X1}
\begin{eqnarray}
{\rm sin}\,\vf \frac{\p \G_0}{\p \vt} + {\rm cot}\,\vt\, {\rm cos}\,\vf
\frac{\p \G_0}{\p \vf} + [\G_0, A_{(1)}] &=& 0 \\
{\rm sin}\,\vf \frac{\p \G_1}{\p \vt} + {\rm cot}\,\vt\, {\rm cos}\,\vf
\frac{\p \G_1}{\p \vf} + [\G_1, A_{(1)}] &=& 0 \\
{\rm sin}\,\vf \frac{\p \G_2}{\p \vt} + {\rm cot}\,\vt\, {\rm cos}\,\vf
\frac{\p \G_2}{\p \vf} + [\G_2, A_{(1)}] -\frac{{\rm cos}\,\vf}
{{\rm sin}^2 \,\vt} \G_3 + \frac{\p A_{(1)}}{\p \vt}&=& 0, \\
{\rm sin}\,\vf \frac{\p \G_3}{\p \vt} + {\rm cot}\,\vt\, {\rm cos}\,\vf
\frac{\p \G_3}{\p \vf} + [\G_3, A_{(1)}] + {\rm cos}\,\vf \G_2 -
{\rm cot}\,\vt\,{\rm sin}\,\vf \G_3 +\frac{\p A_{(1)}}{\p \vf} &=& 0,
\end{eqnarray}
\end{mathletters}
while  the equations $L_{X_{(2)}} \G_j = 0$ read
\begin{mathletters}
\label{X2}
\begin{eqnarray}
-{\rm cos}\,\vf \frac{\p \G_0}{\p \vt} + {\rm cot}\,\vt\, {\rm sin}\,\vf
\frac{\p \G_0}{\p \vf} + [\G_0, A_{(2)}] &=& 0,\\
-{\rm cos}\,\vf \frac{\p \G_1}{\p \vt} + {\rm cot}\,\vt\, {\rm sin}\,\vf
\frac{\p \G_1}{\p \vf} + [\G_1, A_{(2)}] &=& 0, \\
-{\rm cos}\,\vf \frac{\p \G_2}{\p \vt} + {\rm cot}\,\vt\, {\rm sin}\,\vf
\frac{\p \G_2}{\p \vf} + [\G_2, A_{(2)}] -\frac{{\rm sin}\,\vf}
{{\rm sin}^2 \,\vt} \G_3 + \frac{\p A_{(2)}}{\p \vt}&=& 0, \\
-{\rm cos}\,\vf \frac{\p \G_3}{\p \vt} + {\rm cot}\,\vt\, {\rm sin}\,\vf
\frac{\p \G_3}{\p \vf} + [\G_3, A_{(2)}] + {\rm sin}\,\vf \G_2 +
{\rm cot}\,\vt\,{\rm cos}\,\vf \G_3 +\frac{\p A_{(2)}}{\p \vf} &=& 0.
\end{eqnarray}
\end{mathletters}
Finally for $L_{X_{(3)}} \G_j = 0$ we obtain
\begin{equation}
\frac{\p \G_j}{\p \vf} = 0.
\label{X3}
\end{equation}
Here presuppose that $\G_j$ are taken in the form
\begin{eqnarray}
\G_j \,=\,
\left(\begin{array}{cccc}
\G_{j0}^{0}&\G_{j1}^{0}&\G_{j2}^{0}&\G_{j3}^{0}\\[2mm]
\G_{j0}^{1}&\G_{j1}^{1}&\G_{j2}^{1}&\G_{j3}^{1}\\[2mm]
\G_{j0}^{2}&\G_{j1}^{2}&\G_{j2}^{2}&\G_{j3}^{2}\\[2mm]
\G_{j0}^{3}&\G_{j1}^{3}&\G_{j2}^{3}&\G_{j3}^{3}
\end{array}\right),
\label{Gj}
\end{eqnarray}
where the upper indices enumerate the rows. 
From the equation (\ref{X3}) it follows that the $\G_j$'s  are
independent of $\vf$.
Taking into account that the $\G_j$'s are independent of $t$ and $\vf$ we
finally combine the foregoing equations (\ref{X1}) and (\ref{X2})
in the form
\begin{equation}
[\G_0, C] = 0, \quad \frac{\p \G_0}{\p \vt} + [\G_0, D] = 0,
\label{g0}
\end{equation}
\begin{equation}
[\G_1, C] = 0, \quad \frac{\p \G_1}{\p \vt} + [\G_1, D] = 0,
\label{g1}
\end{equation}
\begin{equation}
[\G_2, C] -\frac{1}{{\rm sin}^2 \,\vt} \G_3 + \frac{\p C}{\p \vt}= 0, \quad
\frac{\p \G_2}{\p \vt} + [\G_2, D] + \frac{\p D}{\p \vt}= 0,
\label{g2}
\end{equation}
\begin{equation}
[\G_3, C] + \G_2 + D= 0, \quad
\frac{\p \G_3}{\p \vt} + [\G_3, D] -{\rm cot}\,\vt \G_3 - C =  0,
\label{g3}
\end{equation}
where we define
\begin{eqnarray}
C &=& {\rm cos}\,\vf A_{(1)} + {\rm sin}\,\vf A_{(2)}
= \left(\begin{array}{cc}
0&0\\
0&\tilde{C}\end{array}\right), \quad
\tilde{C} = \left(\begin{array}{cc}
0&1\\
-1/{\rm sin}^2 \,\vt&0
\end{array}\right), \nonumber\\[2mm]
D &=& {\rm sin}\,\vf A_{(1)} - {\rm cos}\,\vf A_{(2)}
= \left(\begin{array}{cc}
0&0\\
0&\tilde{D}\end{array}\right), \quad
\tilde{D} = \left(\begin{array}{cc}
0&0\\
0&- {\rm cot}\,\vt
\end{array}\right). \nonumber
\end{eqnarray}
Solving the equations (\ref{g0} -- \ref{g3}), we find $\G_j$'s which are
independent of $t$ and $\vf$
\begin{eqnarray}
\label{Ggen}
\G_0 &=&
\left(\begin{array}{cccc}
a&\alpha&0&0\\[1mm]
\beta&b&0&0\\[1mm]
0&0&c&-d\,{\rm sin} \,\vt\\[1mm]
0&0&d/{\rm sin} \,\vt&c
\end{array}\right), \quad
\G_1 \,=\,
\left(\begin{array}{cccc}
\gamma&h&0&0\\[1mm]
k&\delta&0&0\\[1mm]
0&0&\mu&-\nu\, {\rm sin} \,\vt\\[1mm]
0&0&\nu/{\rm sin} \,\vt&\mu
\end{array}\right), \nonumber \\  \\
\G_2 &=&
\left(\begin{array}{cccc}
0&0&p& q\,{\rm sin} \,\vt \\[1mm]
0&0&\sigma&\tau\,{\rm sin} \,\vt \\[1mm]
m&\lambda&0&0\\[1mm]
n/{\rm sin} \,\vt &\ve/{\rm sin} \,\vt&0&{\rm cot} \,\vt
\end{array}\right), \quad
\G_3 \,=\,
\left(\begin{array}{cccc}
0&0&-q\,{\rm sin} \,\vt&p\, {\rm sin}^2 \,\vt\\[1mm]
0&0& -\tau {\rm sin} \,\vt &\sigma\,{\rm sin}^2 \,\vt \\[1mm]
-n\,{\rm sin} \,\vt&-\ve {\rm sin}\,\vt&0& -{\rm sin} \,\vt\,
{\rm cos} \,\vt \\[1mm]
m& \lambda&{\rm cot} \,\vt&0
\end{array}\right). \nonumber
\end{eqnarray}

Thus we found the general spherically symmetric anzats
for displacement field $\G$. All the unknown functions in (\ref{Ggen}) 
are the arbitrary functions of $r$ only.

Now the problem is to find these functions as the solutions of the
equations (\ref{cl}).
Taking into account that $\p_t \G_j = 0$ and $\p_{\vf}
\G_j =0$, from (\ref{rel}) and (\ref{H}) for the non-trivial
components of the Riemann tensor we find
\begin{mathletters}
\label{B}
\begin{eqnarray}
B_{10} &=& \frac{\p \G_0}{\p r} + [\G_1,\G_0],\\ 
B_{20} &=& \frac{\p \G_0}{\p \vt} + [\G_2, \G_0],\\ 
B_{30} &=& [\G_3, \G_0],\\ 
B_{12} &=& \frac{\p \G_2}{\p r} - \frac{\p\G_1}{\p \vt} + [\G_1, \G_2],\\ 
B_{13} &=& \frac{\p \G_3}{\p r} + [\G_1, \G_3],\\ 
B_{23} &=& \frac{\p \G_3}{\p \vt} + [\G_2, \G_3].
\end{eqnarray}
\end{mathletters}
Putting (\ref{Ggen}) into (\ref{B}) one can find the non-trivial components 
of the Riemann tensor $B_{ij}$. But we shall not do that since,
for further simplification of our problem we demand the
$\G_j$'s to be invariant under time inversion, i.e., under
$$t \to t'= -t,\quad r \to r' = r,\quad \vt \to \vt'= \vt, \quad
\vf \to \vf' = \vf$$
the $\G_j$'s should remain unaltered. Let us explain from general
point of view what does it mean. Let the transformation $\phi$ on
the manifold $\cM$ maps coordinate patch $U$ onto itself. The
transformation $\phi$ can be represented by smooth functions in $U$
$$\phi: x^i \Longrightarrow \phi^{i} (x); \quad \phi^{-1}: x^i
\Longrightarrow f^{i} (x); \quad \phi^{i} \bigl(f(x)\bigr) = x^i.$$
Under $\phi$\,\,\, $\G$ transforms as follows:
\begin{equation}
\tilde{\G}_{jk}^{\,\,s}(x) = \phi_{l}^{s} \bigl(f(x)\bigr)
\G_{mn}^{\,\,l}\bigl(f(x)\bigr) f_{j}^{m}(x)f_{k}^{n}(x) +
\phi_{l}^{s}\bigl(f(x)\bigr) \p_j f_{k}^{l}(x),
\label{coort}
\end{equation}
where $f_{l}^{s} = \p_l f^s (x), \quad  \phi_{l}^{s} = \p_l \phi^{s}
(x)$. It is said the field with components $\G_{jk}^{\,\,i}$   is
invariant with respect to the transformation $\phi$ if
\begin{equation}
\tilde{\G}_{jk}^{\,\,s}(x) = \phi_{l}^{s} \bigl(f(x)\bigr)
\G_{mn}^{\,\,l}\bigl(f(x)\bigr) f_{j}^{m}(x)f_{k}^{n}(x) +
\phi_{l}^{s}\bigl(f(x)\bigr) \p_j f_{k}^{l}(x) = \G_{jk}^{\,\,s},
\label{coort1}
\end{equation}
At infinitesimal $\phi$, when $f^i (x) = x^i + v^i \epsilon$, from
(\ref{coort1}) it follows that $L_{X} \G_{jl}^{\,\,i} = 0$ where
$X = v^i \frac{\p}{\p x^i}$.

In case of time inversion $f^0 (x) = -t,\, f^1 (x) = r,\,
f^2 (x) = \vt$ and $f^3 (x) = \vf$, hence
$$ F = f_{k}^{l}(x) =
\frac{\p f^{l}(x)}{\p x^{k}} = \left(\begin{array}{cccc}
-1&0&0&0\\[1mm] 0&1&0&0\\[1mm] 0&0&1&0\\[1mm] 0&0&0&1
\end{array}\right).$$
Taking into account that $\p_j f_{k}^{l}(x) = 0$, multiplying
(\ref{coort}) by $f_{s}^{i}(x)$ from the left after a little
manipulation we find the transformation law for $\G_j$'s
\begin{equation}
F \G_j (x) = \bigl[ f_{j}^{m} \G_m(f(x))\bigr] F,
\label{timei}
\end{equation}
or more explicitly
\begin{equation}
F \G_0(x) = -\G_0(f(x)) F,\quad F \G_\mu(x) = \G_\mu(f(x)) F, \quad
\mu = 1,2,3
\label{timeiex}
\end{equation}
From (\ref{timeiex}) we find
\begin{equation}
\G^{\mu}_{0\nu} = 0, \quad \G^0_{00} =0, \quad
\G^0_{\mu1} =\G^0_{\mu2}= \G^0_{\mu3} =0,
\quad \G^1_{\mu0} = \G^2_{\mu0} = \G^3_{\mu0} =0, \quad
\mu,\,\nu =1,2,3. \label{timeinv}
\end{equation}
Thus the $\G_j$'s those are spherically symmetric and invariant
under time inversion:
\begin{eqnarray}
\label{Ggenti}
\G_0 &=&
\left(\begin{array}{cccc}
0&\alpha&0&0\\[1mm]
\beta&0&0&0\\[1mm]
0&0&0&0\\[1mm]
0&0&0&0
\end{array}\right), \quad
\G_1 \,=\,
\left(\begin{array}{cccc}
\gamma&0&0&0\\[1mm]
0&\delta&0&0\\[1mm]
0&0&\mu&-\nu\, {\rm sin} \,\vt\\[1mm]
0&0&\nu/{\rm sin} \,\vt&\mu
\end{array}\right), \nonumber \\  \\
\G_2 &=&
\left(\begin{array}{cccc}
0&0&0&0\\[1mm]
0&0&\sigma&\tau\,{\rm sin} \,\vt \\[1mm]
0&\lambda&0&0\\[1mm]
0&\ve/{\rm sin} \,\vt&0&{\rm cot} \,\vt
\end{array}\right), \quad
\G_3 \,=\,
\left(\begin{array}{cccc}
0&0&0&0\\[1mm]
0&0& -\tau {\rm sin} \,\vt &\sigma\,{\rm sin}^2 \,\vt \\[1mm]
0&-\ve {\rm sin}\,\vt&0& -{\rm sin} \,\vt\,{\rm cos} \,\vt \\[1mm]
0& \lambda&{\rm cot} \,\vt&0
\end{array}\right). \nonumber
\end{eqnarray}
Now putting (\ref{Ggenti}) into (\ref{B}) we obtain the following non-trivial
components of the Riemann tensor
\begin{eqnarray}
\label{riemann1}
B_{10} &=& \left(\begin{array}{cccc}
0&\ba&0&0\\
\bb&0&0&0\\
0&0&0&0\\
0&0&0&0
\end{array}\right), \quad
B_{20} = \left(\begin{array}{cccc}
0&0&-\alpha\,\sigma& -\alpha\,\tau {\rm sin} \,\vt\\
0&0&0&0\\
\lambda\,\beta&0&0&0\\
\ve\,\beta/{\rm sin} \,\vt&0&0&0
\end{array}\right), \nonumber\\[2mm]
B_{30} &=& \left(\begin{array}{cccc}
0&0&-\alpha\,\tau {\rm sin} \,\vt& -\alpha\,\sigma {\rm sin}^2 \,\vt\\
0&0&0&0\\
-\ve\,\beta {\rm sin} \,\vt&0&0&0\\
\lambda\,\beta&0&0&0
\end{array}\right), \quad
B_{12} = \left(\begin{array}{cccc}
0&0&0&0\\
0&0&\bs&\bt {\rm sin} \,\vt\\
0&\bl&0&0\\
0&\be/{\rm sin} \,\vt&0&0
\end{array}\right), \\[2mm]
B_{31} &=& \left(\begin{array}{cccc}
0&0&0&0\\
0&0&\bt {\rm sin} \,\vt&- \bs {\rm sin}^2 \,\vt\\
0& \bs {\rm sin} \,\vt&0&0\\
0&-\bl&0&0
\end{array}\right), \quad
B_{23} = \left(\begin{array}{cccc}
0&0&0&0\\
0&-2 A {\rm sin} \,\vt&0&0\\
0&0& A {\rm sin} \,\vt & B {\rm sin}^2 \,\vt\\
0&0& -B&A {\rm sin} \,\vt
\end{array}\right), \nonumber
\end{eqnarray}
where we define
\begin{eqnarray}
\ba := \alpha^{\pr} - \alpha (\delta - \gamma), \quad
\bb := \beta^{\pr} + \beta (\delta - \gamma),\nonumber\\
\bs := \sigma^{\pr} + \sigma (\delta - \mu) - \tau\,\nu, \quad
\bt := \tau^{\pr} + \tau (\delta - \mu) + \sigma\,\nu,\nonumber\\
\bl := \lambda^{\pr} - \lambda (\delta - \mu) - \ve\,\nu, \quad
\be := \ve^{\pr} - \ve (\delta - \mu) + \lambda\,\nu,\nonumber\\
A := \ve\,\sigma - \tau\,\lambda, \quad
B := \ve\,\tau + \sigma\, \lambda +1.\nonumber
\end{eqnarray}
From (\ref{emt}) we obtain energy-density for the displacement field
$\G_{jk}^{\,\,i}$
\begin{eqnarray}
T_{00} &=& - \ba\,\bb + \frac{2}{r^2} \alpha\,\beta (\sigma\,\lambda +
\tau\, \ve) - \frac{2}{r^2} (\bs\,\bl + \bt\,\be) \nonumber\\
       &-& \frac{2}{r^4} (\lambda\,\tau - \ve\,\sigma)^2 -
 \frac{1}{r^4} \bigl[(\ve\,\sigma - \tau\,\lambda)^2 -
    (\ve\,\tau + \sigma\, \lambda +1)^2\bigr]
\label{too}
\end{eqnarray}

Once the Riemann tensor is defined, we immediately undertake to write the
equations for the functions under consideration. To this end we invoke the
equation (\ref{cl}) that can be rewritten in the form
\begin{equation}
\frac{1}{\sqrt{-g}}\p_i (\sqrt{-g} B^{ij}) + [\G_i, B^{ij}] = 0.
\label{eqmot}
\end{equation}
Here $B^{ij} = B_{pq} g^{ip} g^{jq}$, $\sqrt{-g} = r^2 {\rm sin} \,\vt$ and
$g_{ij} = {\rm diag} (1, -1, -r^2, -r^2 {\rm sin} \,\vt)$.
Inserting (\ref{Ggenti}) and (\ref{riemann1}) into (\ref{eqmot})
we obtain
\begin{mathletters}
\label{b}
\begin{eqnarray}
\ba^{\pr} + \frac{2}{r} \ba + \frac{2}{r^2} (B - 1) \alpha &=& 0,
\label{eba}\\
\bb^{\pr} + \frac{2}{r} \bb + \frac{2}{r^2} (B - 1) \beta &=& 0,
\label{ebb}\\
\alpha \bb - \beta \ba &=& 0, \label{cba} \\
\bs \lambda - \bl \sigma + \bt \ve - \be \tau &=& 0, \label{cep}\\
\bl \tau - \bt \lambda + \bs \ve - \be \sigma &=& 0, \label{cla}\\
\bs^{\pr} - (\mu - \delta) \bs - \alpha \beta \sigma -\bt \nu
+\frac{1}{r^2} (B \sigma + 3 A \tau) &=& 0, \label{ebs}\\
\bt^{\pr} - (\mu - \delta) \bt - \alpha \beta \tau +\bs \nu
+\frac{1}{r^2} (B \tau - 3 A \sigma) &=& 0, \label{ebt}\\
\bl^{\pr} + (\mu - \gamma) \bl - \alpha \beta \lambda -\be \nu
+\frac{1}{r^2} (B \lambda - 3 A \ve) &=& 0, \label{ebl}\\
\be^{\pr} + (\mu - \gamma) \be - \alpha \beta \ve +\bl \nu
+\frac{1}{r^2} (B \ve + 3 A \lambda) &=& 0. \label{ebe}
\end{eqnarray}
\end{mathletters}
The system (\ref{b}) contains ten unknown functions, 
but there is no equation for $\gamma,\,\delta,\,\mu,\,\nu$
those determine $\G_1$.
Let us demand the $\G_j$'s be invariant under space inversion, i.e.,
under
$$t \to t'= t,\quad r \to r' = r,\quad \vt \to \vt'= \pi - \vt, \quad \vf \to
\vf' = \vf$$
the $\G_j$'s should remain unaltered. In this case $F$ in (\ref{timei}) reads
$$ F = f_{k}^{l}(x) = \frac{\p f^{l}(x)}{\p x^{k}} =
\left(\begin{array}{cccc}
1&0&0&0\\[1mm]
0&1&0&0\\[1mm]
0&0&-1&0\\[1mm]
0&0&0&1
\end{array}\right).$$
Hence (\ref{timei}) explicitly reads
\begin{equation}
F \G_2(x) = -\G_2(f(x)) F,\quad F \G_i(x) = \G_i(f(x)) F, \quad i
=0, 1,3 \label{spacex}
\end{equation}
From (\ref{spacex}) we find $\nu
= 0,\, \tau = 0,\, \ve = 0.$ Thus the $\G_j$'s those are spherically
symmetric and invariant under time and space inversion take the form
\begin{eqnarray}
\label{Ggensti}
\G_0 &=&
\left(\begin{array}{cccc}
0&\alpha&0&0\\[1mm]
\beta&0&0&0\\[1mm]
0&0&0&0\\[1mm]
0&0&0&0
\end{array}\right), \quad
\G_1 \,=\,
\left(\begin{array}{cccc}
\gamma&0&0&0\\[1mm]
0&\delta&0&0\\[1mm]
0&0&\mu&0\\[1mm]
0&0&0&\mu
\end{array}\right), \nonumber \\  \\
\G_2 &=&
\left(\begin{array}{cccc}
0&0&0&0\\[1mm]
0&0&\sigma&0 \\[1mm]
0&\lambda&0&0\\[1mm]
0&0&0&{\rm cot} \,\vt
\end{array}\right), \quad
\G_3 \,=\,
\left(\begin{array}{cccc}
0&0&0&0\\[1mm]
0&0&0&\sigma\,{\rm sin}^2 \,\vt \\[1mm]
0&0&0& -{\rm sin} \,\vt\,{\rm cos} \,\vt \\[1mm]
0& \lambda&{\rm cot} \,\vt&0
\end{array}\right). \nonumber
\end{eqnarray}
We again see that $\G_1 \ne 0$. In view of this
let us consider gauge transformations which leave the equation
$L_X \G = 0$ invariant, i.e., find transformations $S$ such that
$L_X \G = 0$ implies $L_X \bG = 0$, where $\bG$ is given by
(\ref{bgm})
$$\bG_i = S \G_i S^{-1} + S \p_i S^{-1}.$$

The natural
choice for the $L_X \G = 0$ to be gauge invariant is to put
\begin{equation} L_X S = 0 \end{equation} or explicitly
\begin{mathletters}
\label{S}
\begin{eqnarray}
{\rm sin}\,\vf \frac{\p S}{\p \vt} + {\rm cot}\,\vt\, {\rm cos}\,\vf
\frac{\p S}{\p \vf} + [S, A_{(1)}] &=& 0
\label{S1}\\
- {\rm cos}\,\vf \frac{\p S}{\p \vt} + {\rm cot}\,\vt\, {\rm sin}\,\vf
\frac{\p S}{\p \vf} + [S, A_{(2)}] &=& 0
\label{S2}\\
\frac{\p S}{\p \vf} &=& 0.
\label{S3}
\end{eqnarray}
\end{mathletters}
In account of (\ref{S3}) we combine (\ref{S1}) and (\ref{S2}) together
to get the equations for determining $S$:
\begin{mathletters}
\label{Sq}
\begin{eqnarray}
\frac{\p S}{\p \vt} -  [S, D] & = & 0  \\
\left[S, C \right] & = & 0.
\end{eqnarray}
\end{mathletters}
General solution of (\ref{Sq}) takes the form
\begin{equation}
S =
\left(\begin{array}{cccc}
\tilde{a}&\tilde{b}&0&0\\[1mm]
\tilde{c}&\tilde{d}&0&0\\[1mm]
0&0&\tilde{e}&-\tilde{f} {\rm sin} \,\vt\\[1mm]
0&0&\tilde{f}/{\rm sin} \,\vt& \tilde{e}
\end{array}\right),
\label{SE}
\end{equation}
with $\tilde{a},\, \tilde{b},\, \tilde{c},\, \tilde{d},\, \tilde{e},\,
\tilde{f}$ being the functions of $r$ only. Now, our assumption of
invariance under space and time inversion leads to the functions
$\tilde{b},\, \tilde{c},\, \tilde{f}$ to be trivial. Hence we obtain the
following expression for $S$:
\begin{equation}
S =
\left(\begin{array}{cccc}
\tilde{a}&0&0&0\\[1mm]
0&\tilde{d}&0&0\\[1mm]
0&0&\tilde{e}&0\\[1mm]
0&0&0&\tilde{e}
\end{array}\right).
\label{SEsti}
\end{equation}
Let us now use the gauge arbitrariness. In doing so we demand $\bG_1$
to be zero.  Then from (\ref{bgm}) , i.e.,
$$\bG_i = S \G_i S^{-1} + S \p_i S^{-1}$$
we obtain equation for fixing gauge
\begin{equation}
\frac{\p S}{\p r} = S \G_1.
\label{Seq}
\end{equation}
that yields the following results
\begin{equation} \tilde{a} = {\rm
exp}\bigl[\int \gamma dr\bigr], \,\, \tilde{d} = {\rm exp}\bigl[\int
\delta dr\bigr] \tilde{e} = {\rm exp}\bigl[\int \mu dr\bigr].
\label{gf}
\end{equation}
Thus, without loss of generality we can put $\G_1 = 0$. Now the system
(\ref{b}) reduces to be
\begin{mathletters}
\label{absl}
\begin{eqnarray}
\alpha^{\pr\pr} + \frac{2}{r} \alpha^{\pr}
+ \frac{2}{r^2}\sigma \lambda \alpha &=& 0, \label{eal}\\
\beta^{\pr\pr} + \frac{2}{r} \beta^{\pr}
+ \frac{2}{r^2}\sigma \lambda \beta &=& 0, \label{ebet}\\
\alpha \beta^{\pr} - \beta \alpha^{\pr} &=& 0, \label{beal}\\
\sigma^{\pr\pr} - \alpha \beta \sigma + \frac{1}{r^2}
(\sigma \lambda + 1) \sigma  &=& 0, \label{sig}\\
\lambda^{\pr\pr} - \alpha \beta \lambda
+ \frac{1}{r^2} (\sigma \lambda + 1) \lambda &=& 0, \label{lam}\\
\lambda \sigma^{\pr} - \sigma \lambda^{\pr} &=& 0. \label{sila}
\end{eqnarray}
\end{mathletters}
From (\ref{beal}) and (\ref{sila}) follow $\beta = c_0 \alpha$ and
$\lambda = d_0 \sigma$, where $c_0$ and $d_0$ are some arbitrary constants.
In this case from (\ref{too}) we find
\begin{equation}
T_{00} = - c_0 \alpha^{pr 2} + \frac{2}{r^2} c_0\,d_0\,\alpha^2\,\sigma^2 -
\frac{2}{r^2} d_0\,\sigma^{\pr 2} + \frac{1}{r^4} (d_0 \sigma^2 + 1)^2.
\label{toost}
\end{equation}
It is obvious from (\ref{toost}) that for the energy to be positive definite
one should simply imply the constants $c_0$ and $d_0$ to be negative, i.e.,
$c_0 < 0$ and $d_0 < 0$.

In spherical coordinates the functions $\alpha,\,\beta,\,\sigma,\,\lambda$
and the constant $d_0$ have the following dimensions:
$[\alpha] = L^{-1},\,\,[\beta] = L^{-1},\,\, [\sigma] = L^{-1},\,\,
[\lambda] = L,\,\, [d_0] = L^{2}$. The constant $c_0$ is dimensionless.

It is obvious that if the system (\ref{absl}) possesses non-trivial
solutions, so does the system (\ref{cl}). One of the special solution
is $\alpha = \alpha_0 /r,\quad  \beta = \beta_0 /r,\quad \lambda = 0$
and $\sigma =0$.

Since the constant $d_0$ is not dimensionless, let us consider the case
when $d_0 = 0$. In other words we assume the function $\lambda$ to be zero.
Under this condition from (\ref{absl}) we find $\alpha = \alpha_0 /r$ and
$\beta = \beta_0 /r$. For $\sigma$ we obtain the equation
\begin{equation}
r^2 \sigma^{\pr\pr} + (1 - \alpha_0 \beta_0) \sigma = 0
\label{harm}
\end{equation}
Introducing a dimensionless parameter $\varrho = r/l$, where $l$ is a
constant such that $[l] = L$, we rewrite the equation (\ref{harm})
\begin{equation}
\varrho^2 \frac{\p^2 \sigma}{\p \varrho^2} + (1 - \alpha_0 \beta_0) \sigma = 0
\label{harm1}
\end{equation}
Defining $b^2 = (1 - \alpha_0 \beta_0)^2 - 1/4$, we find the following
expressions for $\sigma$:
\begin{equation}
\frac{\sigma}{\sqrt{\varrho}} =
\left\{\begin{array}{ccc}
C_1 {\rm cos}\, (b\,{\rm ln}\,\varrho) + C_2 {\rm sin}\,(b\,{\rm ln}\,\varrho),
& b^2 > 0 \\
C_1\, \varrho^b + C_2\, \varrho^{-b}, & b^2 < 0 \\
C_1 + C_2\, {\rm ln}\,\varrho & b^2 = 0
\end{array}\right.
\label{sigmal0}
\end{equation}
where the constants $C_1$ and $C_2$ have the dimension of length.
Thus the system (\ref{absl}) possesses solution and so does the
system (\ref{cl}).

\section{Conclusion}
Summarizing the results obtained we once again underline that
within the framework of gauge symmetry inherent in the concept of
manifold it is natural to consider the linear connection as a gauge field.
Under the gauge symmetry condition it is impossible to demand the condition
$\G_{ij}^{k} = \G_{ji}^{k}$ to be fulfilled, since it is not gauge
invariant. It is shown that the conserving energy-momentum tensor
exists for the displacement field and hence, this field can be
treated within the scope of GR as a material one with deep
geometrical meaning.

To show the similarity of the classical displacement field with the
electromagnetic one and to prove the existence of non-trivial
solutions we have found the static spherically-symmetric anzats.  We
have also shown that its insertion into the equation (\ref{cl})
allows one to obtain the corresponding solutions.

Our conclusion is that together with known long-range interactions
there can exist new type of long-range interactions defined by
displacement field that was the subject of our investigation.

\end{document}